\documentclass[%
 aip,
 amsmath,amssymb,
 reprint,%
]{revtex4-1}

\usepackage{graphicx}
\usepackage{dcolumn}
\usepackage{bm}
\DeclareMathOperator{\sech}{sech}
\makeatletter
\newcommand*{\rom}[1]{\expandafter\@slowromancap\romannumeral #1@}
\makeatother

\usepackage[utf8]{inputenc}
\usepackage[T1]{fontenc}
\usepackage{mathptmx}
\usepackage{xcolor}

\begin{document}
\preprint{AIP/123-QED}

\title{Fast Magnetic Reconnection induced by Resistivity Gradients in 2D Magnetohydrodynamics}
\title[Fast Magnetic Reconnection induced by Resistivity Gradients in 2D Magnetohydrodynamics]{Fast Magnetic Reconnection induced by Resistivity Gradients in 2D Magnetohydrodynamics}

\author{Shan-Chang Lin}
  \email{shan-chang.lin.gr@dartmouth.edu}
\affiliation{Dartmouth College, Hanover, NH 03750}
\author{Yi-Hsin Liu}
\affiliation{Dartmouth College, Hanover, NH 03750}
\author{Xiaocan Li}
\affiliation{Dartmouth College, Hanover, NH 03750}
\date{\today}

\begin{abstract}
Using 2-dimensional (2D) magnetohydrodynamics (MHD) simulations, we show that Petschek-type magnetic reconnection can be induced using a simple resistivity gradient in the reconnection outflow direction, revealing the key ingredient of steady fast reconnection in the collisional limit. We find that the diffusion region self-adjusts its half-length to fit the given gradient scale of resistivity. The induced reconnection x-line and flow stagnation point always reside within the resistivity transition region closer to the higher resistivity end. The opening of one exhaust by this resistivity gradient will lead to the opening of the other exhaust located on the other side of the x-line, within the region of uniform resistivity. Potential applications of this setup to reconnection-based thrusters and solar spicules are discussed. In a separate set of numerical experiments, we explore the maximum plausible reconnection rate using a large and spatially localized resistivity right at the x-line. Interestingly, the resulting current density at the x-line drops significantly so that the normalized reconnection rate remains bounded by the value $\simeq 0.2$, consistent with the theoretical prediction.

\end{abstract}
\maketitle

\section{Introduction}

Magnetic reconnection is a ubiquitous phenomenon in plasma systems that efficiently converts magnetic energy to plasma kinetic energy. In astrophysical environments, observations suggest that magnetic reconnection is the driver of solar flares [e.g., Ref.~\onlinecite{shibata2011solar}] and magnetospheric substorms [e.g., Ref.~\onlinecite{angelopoulos2008tail}]. Particles accelerated by magnetotail reconnection can also be responsible for the generation of aurora borealis and aurora australis  [e.g., Ref.~\onlinecite{birn2012particle} and references therein].  

Sweet-Parker model \citep{parker1957sweet,sweet195814} and Petschek model \citep{petschek1964magnetic} are the two most famous classical reconnection models proposed using the resistive-MHD framework. Reconnection in the Sweet-Parker model develops an elongated diffusion region and has a much smaller reconnection rate compared to that of the Petschek model. While the generalized Sweet-Parker model shows agreements with experiments \citep{ji1998experimental} in the collisional regime, its reconnection rate is many order of magnitude lower in comparison to that inferred by the energy release time-scale of solar flares \citep{parker1963solar}. 

 On the other hand, reconnection in Petschek's model has a short (localized) diffusion region and the outflow is bounded by a pair of standing slow mode shocks. In this reconnection geometry, not only the diffusion region thickness is on the microscopic scale, but also its length. This is in sharp contrast to the system-size long diffusion region length in the Sweet-Parker model. The resulting larger diffusion region aspect ratio corresponds to a faster rate, which is fast enough to explain the time-scale of solar flare observations and geomagnetic substorms. However, numerical simulations show that Petschek reconnection can not form in two-dimensional resistive-MHD simulations if the resistivity is uniform \cite{biskamp1986magnetic,uzdensky2000two}. Such systems appear to lack any mechanism that shortens (localizes) the diffusion region length. 
 To explain this result, Kulsrud\cite{kulsrud2001magnetic} suggested that uniform resistivity can not sustain the magnitude of reconnected (normal to the current sheet) magnetic fields near the end of the diffusion region, which is critical in supporting the open geometry of the Petschek solution.
 Kulsrud further pointed out the importance of the resistivity gradient (in the outflow direction) between the x-line and the end of the diffusion region. The fact that a stable Petschek open geometry can be realized in 2D MHD simulations with a localized resistivity at the x-line supports this idea\cite{ugai1977magnetic}.

In this work, we demonstrate that a stable Petschek-type reconnection can be realized by imposing the resistivity that has a simple one-dimensional (1D) hyperbolic tangent profile varying along the outflow direction. This result is consistent with the finding of Baty et. al.\cite{baty2009petschek}, where only half of a localized resistivity is used in MHD simulations. In this work, we further show that the resistivity gradient in the outflow direction, not in the inflow direction, is the key to inducing Petschek-type reconnection in 2D resistive-MHD. This is also consistent with the result of Yan et. al.\cite{yan1992fast}, where they localized resistivity in the outflow direction only. With a hyperbolic resistivity profile, we can change the transition region length-scale and we find that the reconnection diffusion region will self-adjust its length so that half of the diffusion region just fits into this resistive transition region, not longer nor shorter. The x-line and the flow stagnation point always reside within this transition region near the high resistivity end. This finding further supports Kulsrud's idea \citep{kulsrud2001magnetic}. We also show that the averaged-equation method proposed by Baty et. al.\cite{baty2014formation} can, in certain limits, quantitatively predict the spatial profiles of critical quantities, including the reconnected magnetic field, the outflow speed, and the reconnection layer thickness. Overall, we find that the reconnection rate is determined by both this transition region length and the resistivity value at the x-line. However, if the background resistivity is too high to have a clear separation between the slow shock transitions that bound the outflow exhaust, then the excessive resistive-diffusion therein can reduce the reconnection rate. With this understanding in mind, we go back to study Petschek-type reconnection using a two-dimensional (2D) localized resistivity of different peak strength. Interestingly, we find that the current density ($J$) right at the x-line can drop significantly when a large localized resistivity ($\eta$) is imposed. The resulting maximum plausible normalized reconnection rate ($\eta J$) is around $0.2$, likely being constrained by the force-balance upstream of the diffusion region \citep{yhliu17a}; i.e., the diffusion region physics appears to play a passive role and is forced to match this value.


This paper is organized in the following way. The simulation setup and one of the reconnection simulations with a hyperbolic tangent resistivity profile are described in section \ref{sec:mhd_tanh}. In section \ref{sec:theory}, we explain how resistivity-gradient helps maintain reconnected magnetic field, realizing Petschek-type reconnection, as originally proposed by Kulsrud\cite{kulsrud2001magnetic}. We investigate the details of the resistive-MHD simulations in section \ref{sec:result} by performing a scaling study using the simple hyperbolic tangent resistivity profile where we change the transition region length-scale and the background value of the resistivity. In section \ref{maxrate}, we study the reconnection rates in simulations using exponentially localized resistivity. In particular, we examine how the system responds if one dramatically increases the resistivity right at the x-line. Finally, we summarize and discuss the implication and application of this work in section \ref{discussion}. In the appendix, we briefly summarize the theoretical framework by Baty et. al.\cite{baty2014formation} that predicts key quantities within the reconnection layer; for a given resistivity profile one can solve for the outflow speed, the half-width of the reconnection layer, and the reconnected (normal) magnetic fields. This prediction is compared with our simulation results.
 
\begin{figure*}
\includegraphics[width=\textwidth]{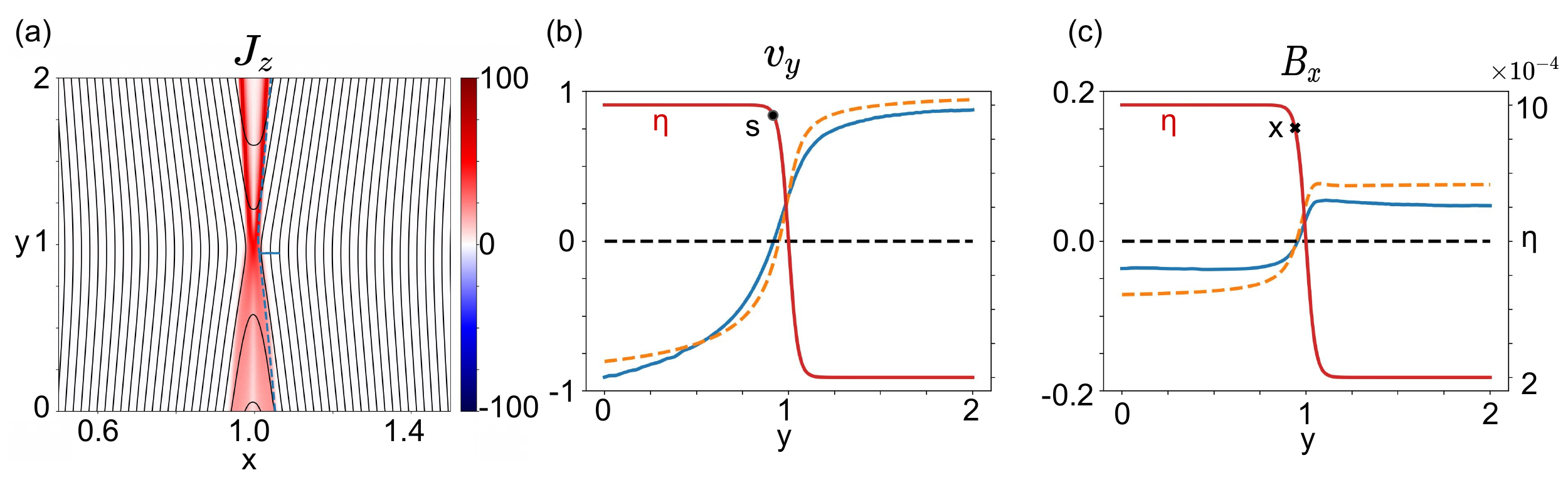}
\caption{(Run $T1$) (a) The current density $J_z$ under a hyperbolic resistivity profile. The contours of the flux function are shown in black. The blue dashed curve marks the layer boundary predicted by the averaged-equations in Appendix \ref{sec:avg}. (b) The outflow speed $v_y$ cut at the symmetric line $x=1$ in simulation (blue solid line) and that predicted by the averaged-equations (orange dashed line). For reference, the resistivity $\eta$ is plotted as the red line. The flow stagnation point is labeled as ``s'' on the $\eta$ profile to show its relative position to the $\eta$-transition region. (c) The normal magnetic fields $B_x$ cut at $x=1$ in simulation (blue solid line) and that predicted by the averaged-equations (orange dashed line). The position of the x-line is labeled as ``x'' on the $\eta$-profile.} \label{fig:1e-3_2e-4}
\end{figure*} 
 
\section{MHD simulation with a hyperbolic tangent resistivity profile}\label{sec:mhd_tanh}

We use Athena\cite{stone2008athena}, a grid-based MHD code, to simulate magnetic reconnection in resistive MHD. The governing equations are
\begin{align}
    &\partial_t\rho+\nabla\cdot(\rho\Vec{v})=0\\
    &\partial_t(\rho\Vec{v})+\nabla\cdot[\rho\Vec{v}\Vec{v}-\Vec{B}\Vec{B}+(p+\frac{B^2}{2})\textbf{I}]=0\\
    &\partial_te+\nabla\cdot[(e+p+\frac{B^2}{2})\Vec{v}-\Vec{B}(\Vec{B}\cdot\Vec{v})]=\nabla\cdot(\Vec{B}\times\eta\Vec{J})\\
    &\partial_t\Vec{B}-\nabla\times[\Vec{v}\times\Vec{B}-\eta\Vec{J}]=0,
\end{align}
where the energy density  $e=p/(\gamma-1)+\rho v^2/2+B^2/2\mu$, and the current density $\Vec{J}=(1/\mu)\nabla\times\Vec{B}$. The ratio of specific heats $\gamma=5/3$ and $\eta$ is the resistivity. The permeability $\mu$ is set to one in code unit. Athena is written based on the finite-volume method (that solves hyperbolic equations) with the higher-order Godunov methods and the constrained transport implemented to ensure the divergence-free condition on the magnetic field. Mass density, momentum, and energy are solved at the center of grid points, while the magnetic field is solved at the center of the grid surfaces.

All simulations in this paper are in 2D. The outflowing (zero-gradient) boundary condition with four ghost cells is used to avoid the saturation due to the flux pileup at outflow regions. Importantly, this boundary better allows the system to evolve into a steady-state. The initial condition is a force-free current sheet, described by
\begin{align}
    \Vec{B}=B_0\tanh{\left(\frac{x-x_c}{\lambda}\right)}\hat{y}+B_0\sech{\left(\frac{x-x_c}{\lambda}\right)}\hat{z},
\end{align}
where $B_0$ is the magnitude of the anti-parallel magnetic field, $\lambda$ is the current sheet half-width and $x_c$ is the x-location of the current sheet. We use $B_0=1$, $\lambda = 0.04$, $x_c=1$, and initial  $\beta=p/(B^2/2\mu)=0.1$ in all simulations and let simulations evolve to quasi-steady states. The Alfv\'en speed based on the reconnecting magnetic field $B_0$ and the background density $n_0=1$ is therefore $V_A=1$ in our unit. The simulation domain in both the x- and y-directions is from 0 to 2. The resolution is $2/512\sim 4\times 10^{-3}$ unless otherwise mentioned. The conclusion discussed in this paper do not change with an initial Harris sheet configuration (not shown). Especially, the reconnection rate in the nonlinear stage is not sensitive to the choice of the force-free current sheets versus Harris sheets.

We apply a hyperbolic tangent resistivity profile varying in the y-direction,
\begin{align}
    \eta_{\tanh}(y)= 0.5\eta_1\left(1 - \tanh{\frac{y-1}{l_{\eta}}}\right) + \eta_2,\label{eq:tanh}
\end{align}
where $l_\eta$ is the resistivity gradient scale. Equation \ref{eq:tanh} gives asymptotic  $\eta=\eta_1+\eta_2$ at $y<1$ (i.e., the lower half-plane in Fig.~\ref{fig:1e-3_2e-4}(a) and other similar figures) and $\eta=\eta_2$ at $y>1$ (i.e., the upper half-plane). Simulations show asymmetric Petschek-type reconnection similar to the results of Baty et. al.\cite{baty2009petschek}, in which they used a localized resistivity profile in the upper half-plane and a uniform resistivity in the lower half-plane. In their simulations, the upper and lower half-planes are connected by a sharp step-like transition.

Figure \ref{fig:1e-3_2e-4} shows a representative run with the hyperbolic tangent $\eta$-profile. The asymptotic resistivity at the upper half-plane is $2\times 10^{-4}$, and that at the lower half-plane is $1\times 10^{-3}$ with the transition length scale $l_{\eta}=0.05$ (i.e., Run $T1$ in Table~\ref{tab:tanh_parameters}). Panel (a) shows the current density in the z-direction $J_z$. The dashed curve in figure 1(a) marks the boundary of one side of the current sheet predicted by the averaged-equations for the given resistivity profile (discussed in Appendix \ref{sec:avg}). Figure 1(b) shows the cut of the outflow speed. Figure 1(c) shows the cut of the reconnected field. Both panels show the simulation results (blue solid curve), those predicted by the averaged-equations (orange dashed curve), and resistivity $\eta$ profiles (red solid curve). The predictions of the averaged-equations agree reasonably well with this run. The deviation of $B_x$ between the averaged-equations solution and the simulation is larger at the lower-half plane, where $\eta$ is larger. We will discuss this effect in section \ref{sec:result}. The reconnection rate of this case is roughly 0.04 in simulation and 0.07 predicted by the averaged-equations.

Interestingly, it appears that the reconnection layer adjusts itself so that the $\eta$-transition region turns to the upper-half of the diffusion region, with both the x-line and flow stagnation point locate near the high-$\eta$ end (see Fig.~\ref{fig:1e-3_2e-4}(b) and (c)) and the upward outflow reaches the plateau at the low-$\eta$ end (see Fig.~\ref{fig:1e-3_2e-4}(b)). Both exhausts are bounded by slow shocks, and the shock transition region is thinner in the upper-half plane because of the lower resistivity, as expected.

\section{The Role of Resistivity Gradient in Petschek model}\label{sec:theory}
Kulsrud\cite{kulsrud2001magnetic} suggested that the reconnected (normal) magnetic fields immediately downstream of the diffusion region are removed by the advection, but can be replenished by the reconnecting magnetic field "rotated" into the normal (x-) direction by resistivity gradient within the diffusion region. For Petschek-type reconnection, if the resistivity is uniform, this advective loss will be higher than the generation if the diffusion length is not on the order of system size. Therefore, the normal magnetic field decreases with time, causing the diffusion region to extend to the system size. If there is a resistivity gradient along the outflow direction, it provides an additional source to generate the normal magnetic field, opening up the outflow geometry. We will briefly discuss the essence of this argument in the following.

If the resistivity is uniform, the x-component of the induction equation (Eq.~(4)) can be expressed as
\begin{align}
    \frac{\partial B_x}{\partial t} = -(\Vec{v}\cdot \nabla) B_x+(\Vec{B}\cdot \nabla) v_x+\eta\left(\frac{\partial^2B_x}{\partial x^2}+\frac{\partial^2B_x}{\partial y^2}\right),\label{eq:ind}
\end{align}
assuming the plasma around the diffusion region is incompressible (i.e., consistent with the simulation results in this paper).
The first term on the right hand side is the down-swiping term which removes the normal magnetic fields at the outflow. Since $v_x$ vanishes along the outflow symmetry line, $-\Vec{v}\cdot \nabla B_x \approx -v_y\partial_yB_x$. The second term on the right hand side is negative in reconnection geometry, thus Kulsrud \cite{kulsrud2001magnetic} removed this term and turned the equal sign ``$=$'' to inequality ``$\leq$'' in Eq.~(\ref{eq:ind}). The resistive term can be approximated as $\eta\partial^2_xB_x=-\eta\partial_x\partial_yB_y$ because $\partial^2_xB_x\gg\partial^2_yB_x$ in a typical reconnection geometry (i.e., low diffusion region aspect ratio) and $\nabla\cdot \Vec{B}=0$ is used. It was further assumed that $B_y =B_0(1-y^2/L^2)$ along the inflow edge of the diffusion region, where $L$ is the system size. By integrating the induction equation from the outflow symmetry line to the inflow edge of the diffusion region, one can obtain
\begin{align}
    a\left\langle\frac{\partial B_x}{\partial t} \right\rangle \leq a\langle -v_y\partial_y B_x \rangle+\frac{2\eta yB_0}{L^2},\label{Kulsred}
\end{align}
where $a(y)$ is the width of the diffusion region. In the steady state, $\partial B_x/\partial t=0$. At the end of the diffusion region, the first term on the right hand side scales as $-a V_AB_x/L'$, where $V_A$ is the Alfv\'enic outflow speed and $L'$ is the half-length of the diffusion region. Consequentially, equation~(\ref{Kulsred}) then gives $-a V_AB_x/L'+2\eta B_0L'/L^2\geq 0$. 
To proceed further, one applies the inflow speed $v_i=(V_A\eta/L')^{1/2}$ of the Sweet-Parker solution \citep{sweet58a,parker1957sweet} based on a diffusion length $L'$, the flux conservation $V_AB_x = v_iB_0$, and the mass conservation $V_A a=v_i L'$ to this inequality,  obtaining $L'\geq L/\sqrt{2}$. This result suggests that a plausible steady-state solution exists only when the diffusion length extend to the system size. 

In contrast, if there is a resistivity gradient, the x-component of the induction equation becomes
\begin{align}
    \frac{\partial B_x}{\partial t} = -(\Vec{v}\cdot \nabla) B_x+(\Vec{B}\cdot \nabla) v_x+\eta\left(\frac{\partial^2B_x}{\partial x^2}+\frac{\partial^2B_x}{\partial y^2}\right)-\notag\\ \frac{\partial\eta}{\partial y}\left(\frac{\partial B_y}{\partial x}-\frac{\partial B_x}{\partial y}\right).
    \label{anomalous}
\end{align}
Since $\partial_x B_y \gg \partial_y B_x$ in the current layer geometry, the last term on the right hand side can be approximated as $\simeq -(\partial_y \eta) (\partial_x B_y)$ and this term is positive if $\eta$ is stronger at the x-line (i.e., $\partial_y \eta < 0$ for $y>1$). Therefore, this resistivity gradient acts as an additional source to supply the normal magnetic field at the outflow, supporting a shorter current sheet length $L'$. The observed open-geometry induced by the simple resistivity gradient in our simulations supports this idea. 

Note that this argument also works with a current-dependent resistivity $\eta (J_z)$ if resistivity can be enhanced by the local current density, i.e., $\partial_{J_z} \eta >0$.
In this situation, using the chain rule the last term of Eq.~(\ref{anomalous}) then becomes $\simeq -(\partial_{J_z}\eta)(\partial_y J_z)(\partial_x B_y)$, that can be positive if the outflows were to be opened up, which requires $\partial_y J_z < 0$ for $y>1$. i.e., this additional supply of $B_x$ can be {\it consistent} with an opening geometry.
This prompted the research on the current-driven anamolous resistivity \citep{papadopoulos1977review,kulsrud2001magnetic, uzdensky2003petschek,drake2003formation,malyshkin2005magnetic}.

Following the argument just discussed above, the $\eta$-transition region in our case is capable of inducing the opening of the upper outflow exhaust. Since the out-of-plane electric field $E_z$ shall be uniform in a 2D steady-state, the fast flux-transport by the opened upper outflow also leads to the opening of the lower outflow exhaust, even though the lower-half plane has a uniform resistivity. The same reason (i.e., uniform $E_z$) might also explain the development of the Petschek-type outflow exhaust (bounded by slow shocks) on the opposite side (respected to an x-line) of a growing plasmoid, that is commonly seen in high-Lundquist number MHD simulations \citep{shibayama2015fast}; i.e., the growth of a plasmoid opens up the outflow on one side of the x-line, the outflow exhaust on the other side consequently develops open geometry as well.



\section{Scaling Study using MHD Simulations}
In this section, we performed a systematic numerical study of magnetic reconnection using the hyperbolic tangent resistivity profile specified in Eq.~(\ref{eq:tanh}), then we conduct a separate set of study to explore the maximum plausible reconnection rate using a spatially localized exponential profile at the x-line, as specified in Eq.~(\ref{eq:exp}). 
\begin{table}[h]
    \centering
    \caption{Simulation parameters with a resistivity of hyperbolic tangent profile. $\eta_{bottom}$ denotes the resistivity value at the lower half-plane ($y<1$), and $\eta_{top}$ denotes the resistivity value at the upper half-plane ($y>1$). $l_\eta$ is the resistivity scale length, and $\beta=p/(B^2/2\mu)$ is the initial upstream plasma beta.}
    \hspace*{-1.5cm}
    \begin{tabular}{c|c|c|c|c}
   \hline
    \hline
        Run & $\eta_{bottom}=\eta_1+\eta_2$ & $\eta_{top}=\eta_2$ & $l_{\eta}$ & $\beta$ \\
        \hline
         $T1$ & $1\times 10^{-3}$ & $2\times 10^{-4}$ & $0.05$ & 0.1\\
         $T2$ & $1\times 10^{-3}$ & $2\times 10^{-4}$ & $0.1$ & 0.1\\
         $T3$ & $1\times 10^{-3}$ & $2\times 10^{-4}$ & $0.2$ & 0.1\\
         $T4$ & $1\times 10^{-3}$ & $2\times 10^{-4}$ & $0.4$ & 0.1\\
         $T5$ & $2\times 10^{-3}$ & $1.2\times 10^{-3}$ & $0.05$ & 0.1\\
         $T6$ & $5\times 10^{-3}$ & $4.2\times 10^{-3}$ & $0.05$ & 0.1\\
         $U1$& $1\times 10^{-3}$ & $1\times 10^{-3}$ & NA& 0.1\\
         \hline
         \hline
    \end{tabular}
    
    \label{tab:tanh_parameters}
\end{table}

\subsection{With a hyperbolic tangent $\eta$ profile}\label{sec:result}


\begin{figure}
    \centering
    \includegraphics[width=\linewidth]{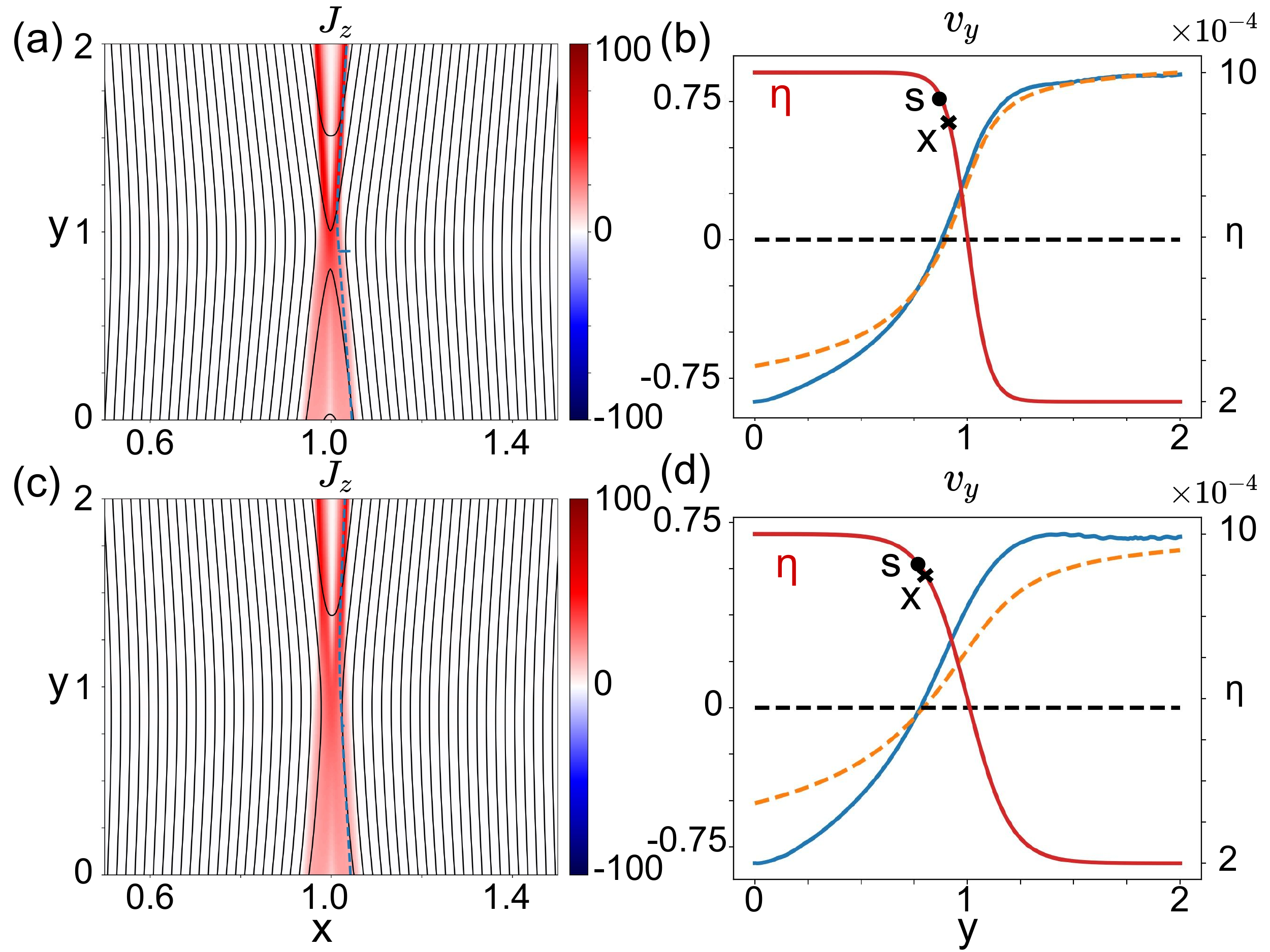}
    \caption{(Runs $T2$ and $T3$) Panels (a) and (b) show the current density $J_z$ and the outflow speed $v_y$ cut at $x=1$ at the nonlinear state (blue from simulation, orange from theory) of Run $T2$. Panels (c) and (d) are for Run $T3$. These two runs are only different in the gradient scale $l_{\eta}$, as also illustrated by the red curves in (b) and (d). Note that the solutions of the averaged-equations (orange) deviate from simulations (blue) more for larger $l_\eta$, that is, smaller resistivity gradient. The locations of the stagnation point and x-line for both cases are labeled as ``s'' and `x'' on the $\eta$-profiles in panels (b) and (d).}
    \label{fig:rs}
\end{figure}

The system with a hyperbolic tangent resistivity profile (Eq.~(\ref{eq:tanh})) tends to evolve into a state where the upper half portion of the diffusion region adjusts itself to the gradient scale length $l_{\eta}$ of the given hyperbolic tangent $\eta$ profile. This observation becomes clearer when one varies $l_{\eta}$.
Figures \ref{fig:rs} shows the current density $J_z$ and the outflow speed $v_y$ cut at $x=1$ for $l_{\eta}=0.1, 0.2$ (Run $T2$ and $T3$) at $t=10$, after reaching quasi-steady states. Combining with the result of Run $T1$ with $l_{\eta}=0.05$ in Fig.~\ref{fig:1e-3_2e-4}, we conclude that the x-line and the stagnation point are both located near the high-$\eta$ end within the transition region. However, these two points do not coincide with each other due to the outflow asymmetry introduced by this resistivity profile, and the stagnation point is always closer to the high-$\eta$ plateau compared to the x-line. In the positive y-direction, outflow speed reaches the Alfv\'enic plateau near the low-$\eta$ end.

\begin{figure}
    \centering
    \includegraphics[width=.5\textwidth]{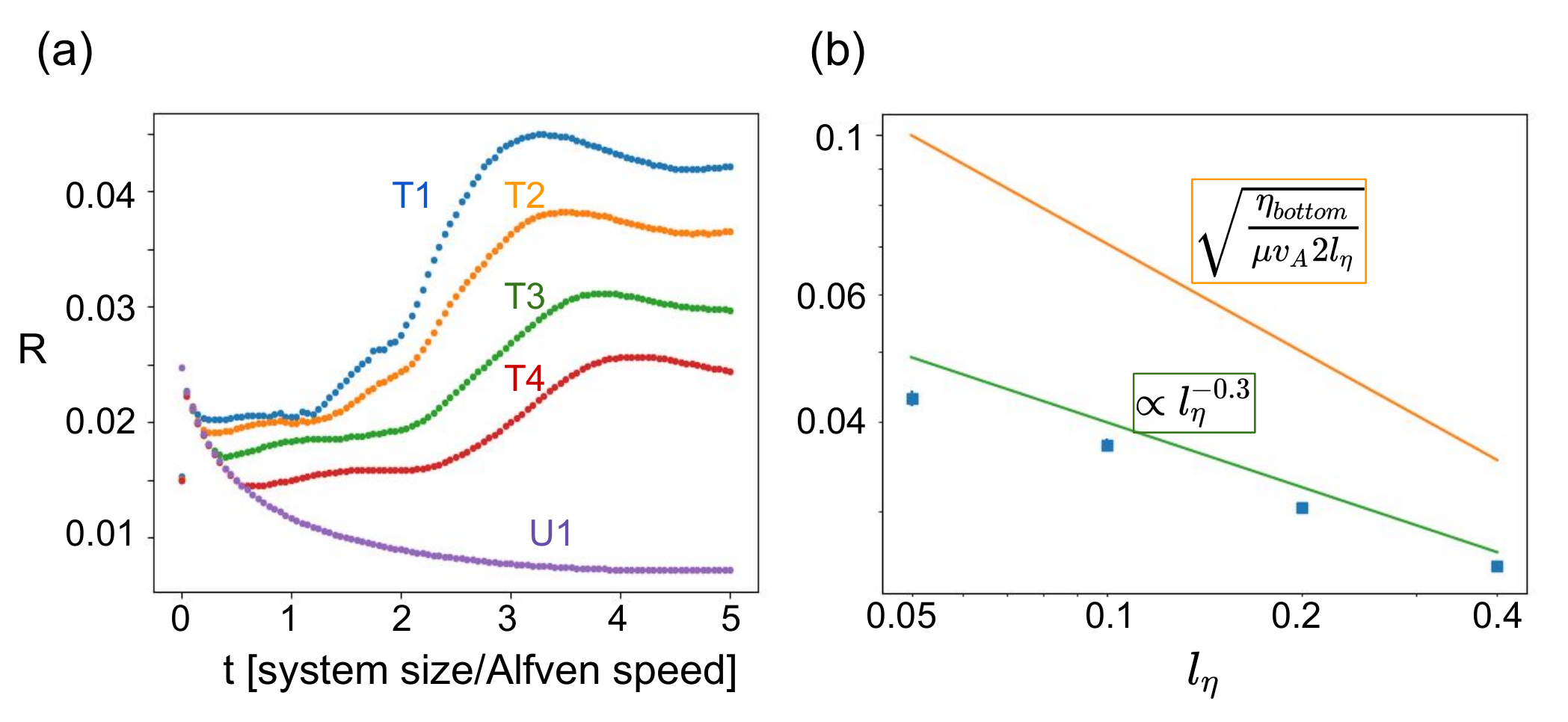}
    \caption{(a) The time evolution of normalized reconnection rates of Runs $T1$, $T2$, $T3$, and $T4$, which have different resistivity gradient scale $l_{\eta}$. A uniform resistivity $\eta=1\times 10^{-3}$ case ($U1$) is also plotted for comparison. Reconnection rates decrease as $l_\eta$ increases. The initial drops are due to the current sheets broadening that reduces $J_z$ at the x-line. (b) Reconnection rates versus the gradient scale length. Blue dots are the average reconnection rates after they reach the peak values. Equation (\ref{eq:rates}) is plotted as the orange line, and a scaling $R \propto l_\eta^{-0.3}$ is plotted as the green line for comparison.}
    \label{fig:rs_rate}
\end{figure}

Reconnection rates in these simulations are calculated using the out-of-plane electric field $E_{z,xline} = \eta_{xline} J_{z,xline}$ right at the x-line, and they are normalized as $R \equiv E_{z,xline}/(B_0V_A)$,
where $B_0$ is the asymptotic value of the magnetic field at the upstream region, and $V_A = B_0/(\mu\rho)^{1/2}$ is the Alfv\'en speed calculated using this upstream magnetic field and density. The evolution of reconnection rates for different $l_{\eta}$ are shown in figure \ref{fig:rs_rate}(a). A uniform resistivity $\eta=1\times 10^{-3}$ case ($U1$) is also plotted for comparison. The reconnection rates decrease as $l_{\eta}$ increases, and it can be explained by the following simple analysis;
since the normalized rate $R =\eta_{xline}J_{z,xline}/(B_0V_A)$ and we know $R\simeq \delta/L'$ from the Sweet-Parker scaling \citep{parker1957sweet}, where $\delta$ and $L'$ is the half-thickness and the length of the ``upper'' diffusion region (i.e., $y > y_{xline}$), respectively. On the other hand, our simulation demonstrates that $L'\simeq 2l_\eta$ and $\eta_{xline}\simeq \eta_{bottom}$. And, we can approximate $J_{z,xline}\simeq B_0/(\mu\delta)$ in the small diffusion region aspect ratio ($\delta/L'$) limit. Combining all these relations, one can derive the scaling of the reconnection rate to be

\begin{equation}
R\simeq \sqrt{\frac{\eta_{xline}}{\mu V_A 2l_\eta}}\simeq \sqrt{\frac{\eta_{bottom}}{\mu V_A 2l_\eta}}.\label{eq:rates}
\end{equation}
 The reconnection rate is thus determined by both the resistivity gradient length and the resistivity value right at the x-line, which is close to $\eta_{bottom}$ in these runs. Figure \ref{fig:rs_rate}(b) shows the reconnection rates, averaged after reaching the peak, versus the resistivity gradient scale length $l_\eta$ in a log-log scale plot. The simulation results are shown as blue dots and the prediction from equation (\ref{eq:rates}) is plotted as the orange line. $R\propto l_\eta^{-0.3}$ is also plotted (green line) for comparison, and one can see that the measured reconnection rates compare better with $l_\eta^{-0.3}$ scaling, instead of $l_\eta^{-0.5}$. This could be caused by the fact that the locations of the x-lines are not exactly at the edge where resistivity starts to decrease (see Fig. \ref{fig:rs}). There is also about a factor of two difference that is not captured by this simple scaling, but Eq.~(\ref{eq:rates}) does qualitatively explain the decreasing trend.

\begin{figure}
    \includegraphics[width=\linewidth]{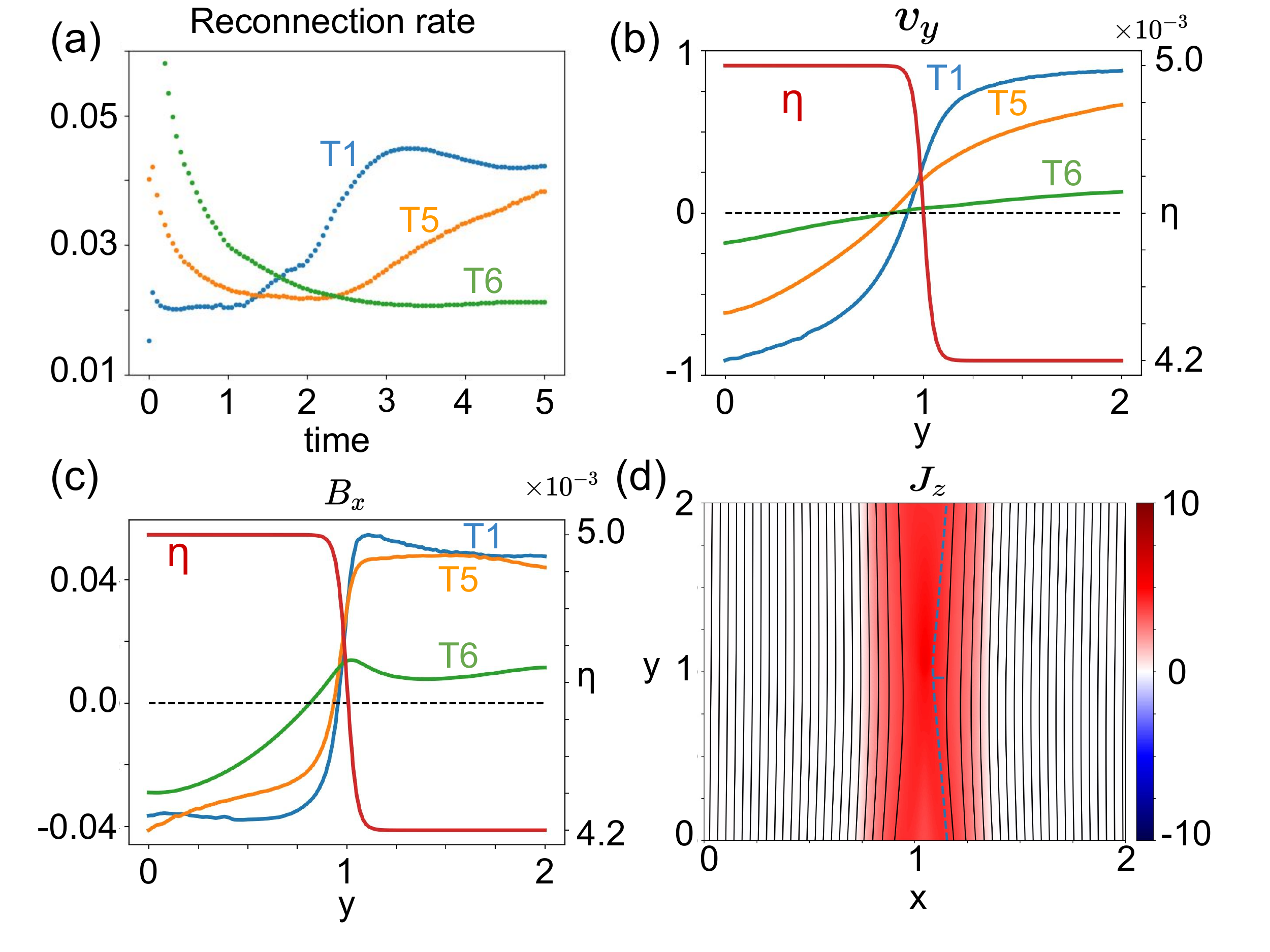}
    \caption{(Run $T1$, $T5$ and $T6$) (a) Reconnection rates with the same resistivity gradient (determined by $\eta_1=8\times 10^{-4}$ and $l_{\eta}=0.05$) but with a different background resistivity $\eta_2$ from low to high in Runs $T1$ (blue), $T5$ (orange), and $T6$ (green). (b) The outflow speed $v_y$ cut at $x=1$ at the nonlinear state. (c) The normal magnetic field $B_x$ cut at $x=1$ at the nonlinear state. In panels (b) and (c), the resistivity $\eta$ of run $T6$ is shown as the red line. (d) The current density $J_z$  of run $T6$ at the nonlinear state. While the flux function is slightly bent inside the diffusion region, the rather thick current sheet extend to the outflow boundaries.}
    \label{fig:tanheta5e-3}
\end{figure}

In the following, we also investigate the effect of background resistivity, that is parametrized by $\eta_2$  of the hyperbolic tangent profile; we increase $\eta_2$ while keeping the same $\eta_1$ and $l_\eta$. While the solutions from the averaged-equations (Appendix A) suggest the increase of reconnection rate, our simulations show an opposite trend. This is demonstrated by the Runs in Fig.~\ref{fig:tanheta5e-3}(a) where $\eta_1=8\times 10^{-4}$, but Run $T1$ has $\eta_2=2\times 10^{-4}$, Run $T5$ has $ \eta_2=12\times 10^{-4}$, and Run $T6$ has $\eta_2=42\times 10^{-4}$. The reconnection rates decrease as the background resistivity increases.

Similar to the thickness of the reconnection diffusion region, the shock thickness also increases with the background resistivity. This could introduce a finite current density $J_z$ within the entire outflow exhaust, as seen in Fig.~\ref{fig:tanheta5e-3}(d).
According to Ohm's law
\begin{align}
    E_z=v_y B_x+\eta J_z
    \label{Ohms}
\end{align}
since $E_z$ along the outflow symmetry line should be uniform in the steady state. If the current density $J_z$ remains significant at the outflow region and the reconnection electric field $E_z$ does not increase with a higher background $\eta$, then $v_y B_x$ should be smaller; this is consistent with the significant drop of $B_x$ and $v_y$ observed in Fig. \ref{fig:tanheta5e-3}(b) and (c). 

\begin{figure}
    \centering
    \includegraphics[width=.5\linewidth]{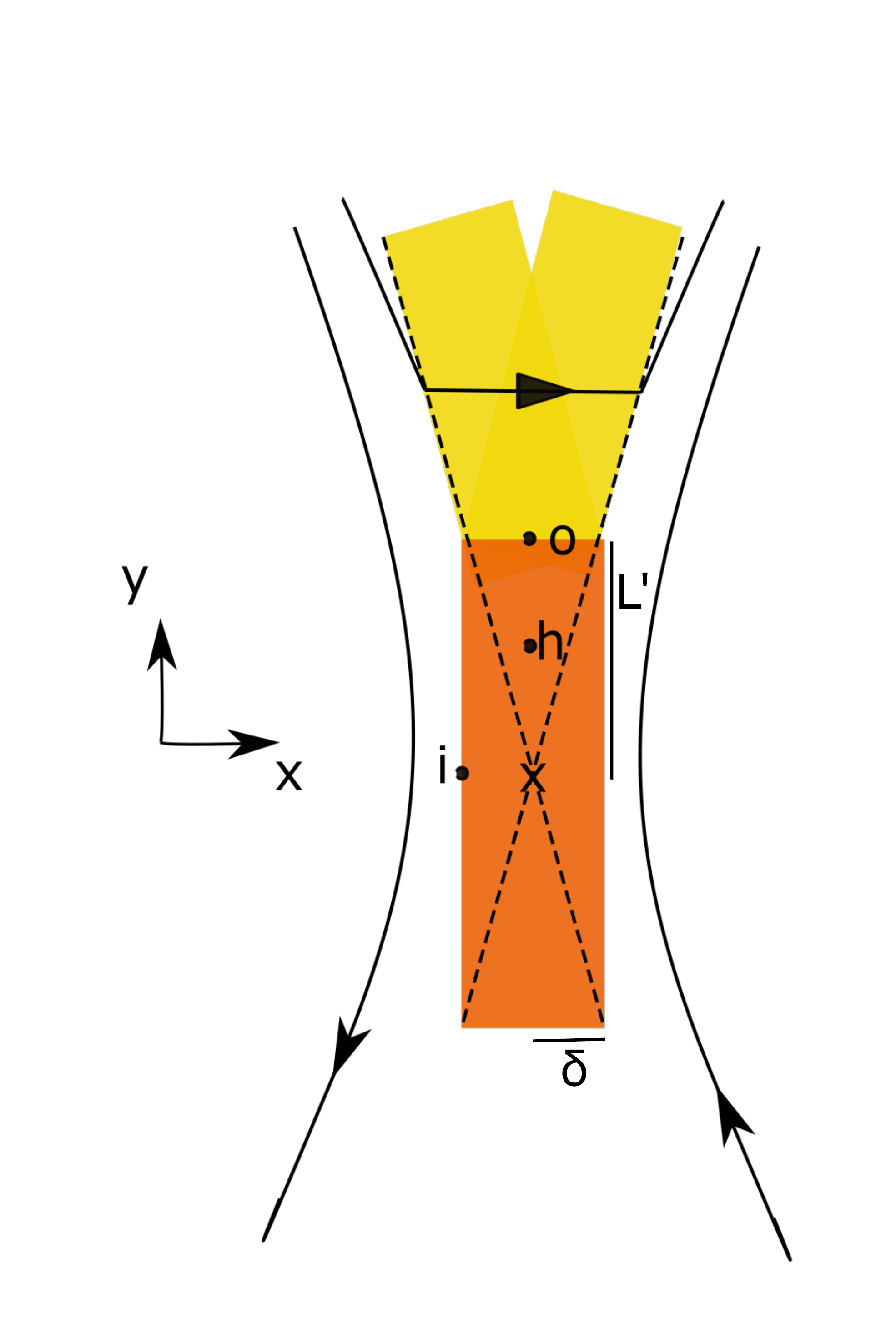}
    \caption{The orange rectangle represents the diffusion region. The diffusion region length ($L'$) is determined by the imposed resistivity gradient scale ($l_{\eta}$). The yellow region illustrates the thickness of downstream slow shock transitions. Point {\bf o} locates at the end of the diffusion region. Point {\bf i} is at the inflow edge of the diffusion region. Point {\bf h} is halfway between the x-line and point {\bf o}. The thickening of the shock transition regions due to a large background resistivity widens the coverage of the non-ideal region, immersing point {\bf o} with a finite $\eta J_z$.}
    \label{fig:diffusion}
\end{figure}

\begin{figure*}
    \centering
       \includegraphics[width=.9\linewidth]{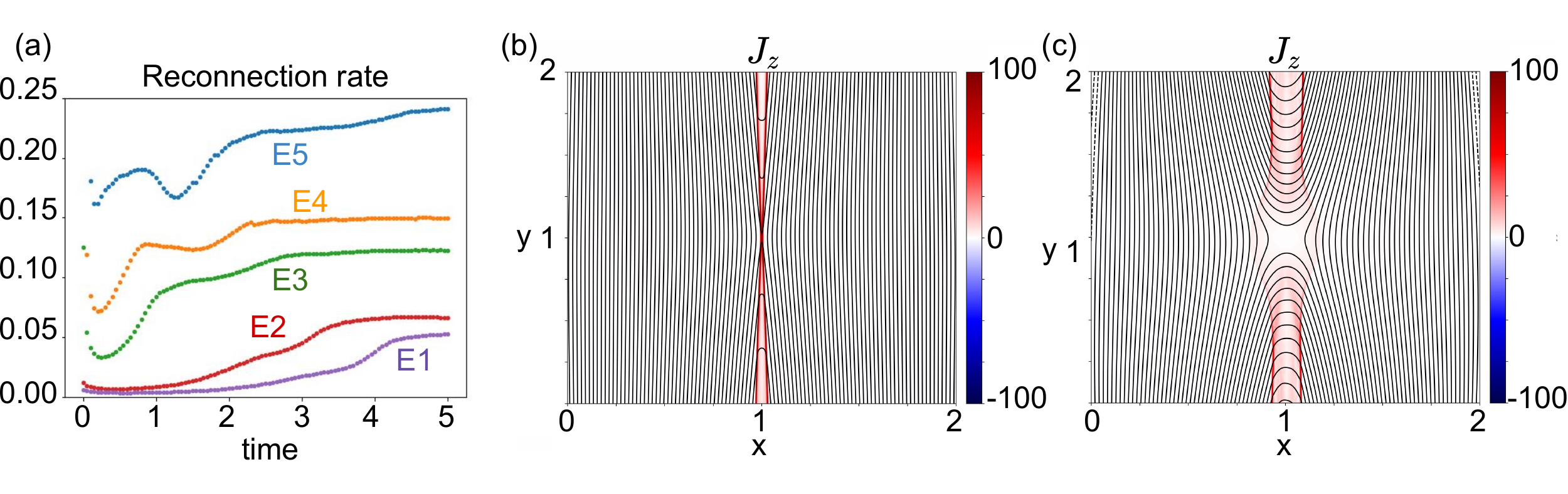}
    \caption{Panel (a) shows the scaling of reconnection rates in simulations with a spatially localized resistivity at the x-line, including Runs $E1$ (purple), $E2$ (red), $E3$ (green), $E4$ (orange), and $E_5$ (blue). Panels (b) and (c) show the current density and flux function of Run $E1$ and Run $E5$, respectively, at late time. Note that the current density drops significantly at the x-line for an extremely strong resistivity in panel (c), so that $\eta J_z$ remains bounded.}
    \label{fig:max_rate}
\end{figure*}

In addition, the current sheet tends to extend to the outflow boundary if the entire reconnection layer is immersed within the non-ideal region with a finite non-ideal electric field $\eta J_z$. To illustrate the underlying reason, we will use the notations in Figure \ref{fig:diffusion}. The inflow and outflow quantities are evaluated at point {\bf i} and {\bf o}, respectively. Point {\bf h} is at the middle between the x-line and point {\bf o}. $L'$ and $\delta$ are the half-length and the half-width of the diffusion region. Near the inflow region the out-of-plane electric field at point {\bf i} matches the value at the x-line, $E_{z,i}=v_{x,i}B_{y,i}\simeq E_{z,xline}$. At point {\bf o}, we have $E_{z,o}=v_{y,o}B_{x,o}+\eta J_{z,o}$ from Ohm's law (Eq.~(\ref{Ohms})). 
Using the Maxwell-Faraday's law $\nabla\times{\Vec{E}}=-\partial_t\Vec{B}$ and the flux conservation $v_{y,o}B_{x,o}\simeq v_{x,i}B_{y,i}$, we can estimate the time derivative of the reconnected field at point {\bf h} to be $\partial_tB_{x,h}\simeq -\partial_y E_z\simeq -(E_{z,o}-E_{z,xline})/L'\simeq -(v_{y,o}B_{x,o}+\eta J_{z,o}-v_{x,i}B_{y,i})/L'=-\eta J_{z,o}/L'$. Therefore, the time derivative of the reconnected field at point {\bf h} is negative ($\partial_tB_{x,h}<0$) if the current density at point {\bf o} is finite ($J_{z,o}>0$); a situation that occurs when there is no clear separation between the pair of slow shocks at the outflow region. The reconnected magnetic field $B_{x,h}$ at point {\bf h} thus tends to decrease with time and the current layer will extend to the boundary, reducing the opening geometry as seen in Fig.~\ref{fig:tanheta5e-3}(d). Note that this mechanism is different from Kulsrud's idea discussed in section \ref{sec:theory}. In short, the wide coverage of the non-ideal region (i.e., the thickening of the shock transition region) due to a large background resistivity causes excessive resistive-diffusion, resulting in an extended current sheets despite the imposed sharp resistivity gradient.

Finally, we also performed simulations (not shown) with a hyperbolic tangent resistivity profile that varies along a direction at a $45^{\circ}$ angle from the outflow ($y$-) direction and found that Petschek-type reconnection can still be realized. This further suggests that the resistivity gradient projected in the outflow direction is sufficient in facilitating open outflow geometry.


\subsection{With a spatially localized $\eta$ at the x-line}\label{maxrate}

In this sub-section, we study how the system responds to an extremely large resistivity spatially localized around the x-line. For the same reason discussed before, Petschek-type reconnection will be realized because of the presence of resistivity gradient along the outflow direction. Special attention is dedicated here to find the maximum plausible reconnection rate potentially applicable to all reconnection systems. To do so, we adopt the following $\eta$-profile that exponentially decays out of the center of the simulation domain (i.e, the x-line).  
 

\begin{align}
    \eta_{\exp}=\eta_1\exp\left( -\frac{r}{l_{\eta}}\right) + \eta_2,
    \label{eq:exp}
\end{align}
where the distance to the center is parameterized by the radius $r\equiv[(x-1)^2+(y-1)^2]^{1/2}$.
The peak value of the resistivity is $\eta_{xline}=\eta_1+\eta_2$ at the center $(x,y)=(1,1)$ and the lowest value is $\eta_2$ in the background. Figure~\ref{fig:max_rate}(a) shows the reconnection rates of different $\eta_{xline}=\eta_1+\eta_2$ from $5\times 10^{-4}$ to $1$. The parameters of the simulations are summarized in table \ref{tab:sim_parameters_exp}.

\begin{table}
    \centering
    \caption{Simulation parameters with a resistivity of exponential profile.}
    \hspace*{-1.5cm}
    \begin{tabular}{c|c|c|c|c}
   \hline
    \hline
        Run & $\eta_{xline}=\eta_1+\eta_2$ & $\eta_{background}=\eta_2$ & $l_{\eta}$ & $\beta$ \\
        \hline
         $E1$ & $5\times 10^{-4}$ & $1\times 10^{-4}$ & $0.05$ & 0.1\\
         $E2$ & $1\times 10^{-3}$ & $1\times 10^{-4}$ & $0.05$ & 0.1\\
         $E3$ & $1 \times 10^{-2}$ & $1\times 10^{-4}$ & $0.05$ & 0.1\\
         $E4$ & $5 \times 10^{-2}$ & $1\times 10^{-4}$ & $0.05$ & 0.1\\
         $E5$ & $1$ & $1\times 10^{-4}$ & $0.05$ & 0.1\\
         \hline
         \hline
    \end{tabular}
    
    \label{tab:sim_parameters_exp}
\end{table}

From Fig.~\ref{fig:max_rate}(a), it is clear that, with a fixed resistivity gradient scale $l_\eta=0.05$, a larger $\eta_{xline}$ results in a higher reconnection rate.
Panel (b) shows the current density of Run $E1$, which has a well-localized diffusion region, and the open outflow exhausts are bounded by the sharp transitions of slow shocks. The reconnection rate of this case ($\simeq 0.04$) is the lowest in panel (a) because it has the lowest $\eta_{xline}=5 \times 10^{-4}$.
Run $E2$ has a higher resistivity $\eta_{xline}=1\times 10^{-3}$. Its rate is shown in red in Fig.~(\ref{fig:max_rate})(a), which is very close to the rate measured in hyperbolic tangent resistivity simulation that has the same $\eta_1+\eta_2$ and $l_{\eta}$ (Run $T1$, recall that the x-line develops close to the high-$\eta$ end, thus $\eta_{xline}\simeq \eta_{bottom}= \eta_1+\eta_2=1\times 10^{-3}$ in this run). This is consistent with our expectation that the resistivity gradient scale and strength right at the x-line determined the rate (if the background resistivity is low enough).

The most surprising case is Run $E5$ that has an extremely strong $\eta_{xline}= 1.0$, that is $2000$ times larger than that in Run $E1$ (Fig.~\ref{fig:max_rate}(b)). In this case, the current sheet broadens and the dramatic drop of the current density right at the x-line is a pronounced feature, as shown in Fig.~\ref{fig:max_rate}(c). Consequently, the reconnection rate $E_R=\eta_{xline} J_{z,xline}$ remains on the order of the typical fast rate $0.1$\footnote{We learned in private conversations that Judit P\'erez-Coll and Michael Hesse also get the same result independently. Their study will be published in the future.}
, as shown in Fig.~\ref{fig:max_rate}(a). This numerical experiment demonstrates that the reconnection rate is bounded by physics outside of the diffusion region, presumably by the force-balance in the upstream region\cite{yhliu17a}. No matter how strong and localized the resistivity is, the diffusion region is forced to adjust itself to accommodate this maximum plausible rate $\simeq 0.2$. The slight increase of the rate at $E5$ near the end comes from the numerical effects at the boundary, which we will leave for future investigation.

Along this line of discussion, it is also interesting to compare the maximum reconnection rate predicted by Petschek\cite{petschek1964magnetic} 
\begin{align}
R_\text{Petschek}\simeq \frac{\pi}{8\ln{S}},
\label{Petschek}
\end{align}
where $S= LV_A/\eta$ is the Lundquist number, $L=1$ is system size, $V_A$ is the Alf\'ven speed and $\eta$ is the resistivity. For Run $E5$ the relevant $\eta$ is the value near the x-line, $\eta_{xline}=1$, because in Petschek's derivation $\eta$ is introduced by matching the reconnection electric field at the x-line to the value immediately upstream of the diffusion region; i.e., $E_{z,xline}=\eta_{xline}J_{z,xline}\simeq v_{x,i}B_{y,i}$ (the same notations used in the discussion of Fig.~\ref{fig:diffusion}). Thus, the reconnection rate is predicted to approach infinity in Petscheck's model; i.e., $1/\mbox{ln}(1\times 1/1)\rightarrow \infty$ in Eq.~(\ref{Petschek}). However, the reconnection rate is still on the order of the typical fast rate value $0.1$.
This discrepancy likely also results from the lack of a self-consistent consideration of the force-balance upstream of the diffusion region, that applies to all reconnection systems \cite{yhliu17a}, either collisional or colliionless.


\section{Summary and Discussion}\label{discussion}

In summary, we find that steady Petschek-type fast magnetic reconnection can be generated as long as there is an $\eta$-gradient along the reconnection outflow direction in MHD simulations. This finding supports the idea of Kulsrud\cite{kulsrud2001magnetic} that suggested a resistivity gradient can provide additional supply of the normal magnetic fields within the diffusion region, balancing the loss by the outflow convection. In simulations with a resistivity that has a simple hyperbolic tangent profile, the opening exhaust on one side leads to the opening on the other side because the electric field is uniform in a 2D steady-state. The diffusion region self-adjusts its half-length to fit the resistivity gradient length. Therefore, increasing the resistivity gradient length will decrease the reconnection rate. 


The solutions of the averaged-equations (Appendix A) proposed in Refs.~\onlinecite{seaton2009analytical,baty2014formation} show reasonable agreement with the hyperbolic tangent resistivity simulations when the resistivity gradient is large and the resistivity background is small. The solutions do not agree well with our simulations when there is a large background resistivity or a small resistivity gradient, and we provide an explanation to address the effect of large background resistivity. Besides, the reconnection rates predicted by the averaged-equations are not bounded in the large background $\eta$ limit. It is likely due to the lack of consideration on the upstream force-balance, which is critical in limiting the reconnection rate \citep{yhliu17a}.


This work demonstrates that anti-parallel magnetic fields that thread these two regions will prefer to reconnect at this interface, where the energy release is most efficient. The fact that we can induce fast reconnection in collisional plasmas using resistivity gradient, and confine the x-line within the transition region, may be handy for the design of reconnection-based thrusters \citep{bathgate2018thruster}; i.e., a collisional plasma might be more accessible than collisionless plasma in compact devices, and we know how to realize a stable single x-line fast reconnection.

In natural plasmas, a resistivity gradient can arise at the sharp transition layer of temperature and density, or at the interface between different ion species\citep{cohen1950electrical,spitzer1953transport}, such as the solar transition region\cite{roussev2001modelling} or the photosphere-chromosphere interface\cite{litvinenko1999magnetic}. The resistivity gradient scale at the interface between the photosphere and chromosphere is estimated as 100 km \cite{ni2020magnetic}, which is smaller than (or, at least, not larger than) the size of flux tubes observed\cite{litvinenko1999magnetic}($\approx$200-300 km, which will be the system size of our simulations). This suggests that our result could be relevant to reconnection phenomena occurring in the lower solar atmosphere. In particular, our work predicts that solar spicules\cite{moore2011solar,samanta2019generation}, if driven by reconnection, may tend to develop at the altitude of a sharp resistivity gradient
.

On a separate issue, although resistive-MHD is often deemed inadequate to address the physics at a diffusion-region scale, it nevertheless allows us to test out the maximum plausible reconnection rate in a clean fashion. Specifically, we can control the strength and localization of resistivity, which simply cannot be done in fully kinetic simulations; i.e., fully kinetic simulations generate dissipation and diffusion self-consistently \citep{hesse11a}. We found that the reconnection rates are well-bounded by value $\simeq 0.2$, no matter how strong the localized resistivity is. The existence of this upper bound can be explained by the upstream force-balance in the MHD region \citep{yhliu17a}. This fact has a significant implication, suggesting that a strong anomalous resistivity will not further increase much the typical fast rate of order 0.1 reported in 2D laminar kinetic simulations \citep{birn01a,shay98a}. Notably, the reconnection rates observed by NASA's Magnetospheric Multiscale (MMS) mission \cite{burch2020electron,torbert18a,RNakamura18a,TKMNakamura18a,genestreti18b} are consistently bounded by the maximum plausible value 0.2 demonstrated here.  

In conclusion,  this work shows that a resistivity gradient can efficiently induce a spatially localized diffusion region and fast reconnection in collisional plasmas. We further expect that if the local reconnection electric field excesses the Dreicer runaway value, the diffusion region plasma transitions to the collisionless regime \cite{stanier19a,daughton09a}, and kinetic physics \cite{birn01a, shay98a, cassak05a, rogers01a,yhliu14a, yamada2011mechanisms, malyshkin2009model, andres2016influence, hesse11a} can take over to continue fast reconnection. 

\acknowledgments
We gratefully acknowledge helpful discussions with Michael Hesse, Judit P\'erez-Coll, Terry Forbes, Jongsoo Yoo, Jim Klimchuk, and Chengcai Shen. Contributions from S.L., Y.L. and X.L. are based upon work funded by the National Science Foundation Grant No. PHY-1902867 through the NSF/DOE Partnership in Basic Plasma Science and Engineering and NASA MMS 80NSSC18K0289. 

\section*{Data Availability}
Raw data were generated at the NERSC Advanced Supercomputing large scale facility. Derived data supporting the
findings of this study are available from the corresponding author upon reasonable request

\appendix
\section{Averaged MHD Equations}\label{sec:avg}

\begin{figure*}
    \centering
    \includegraphics[width=.25\linewidth]{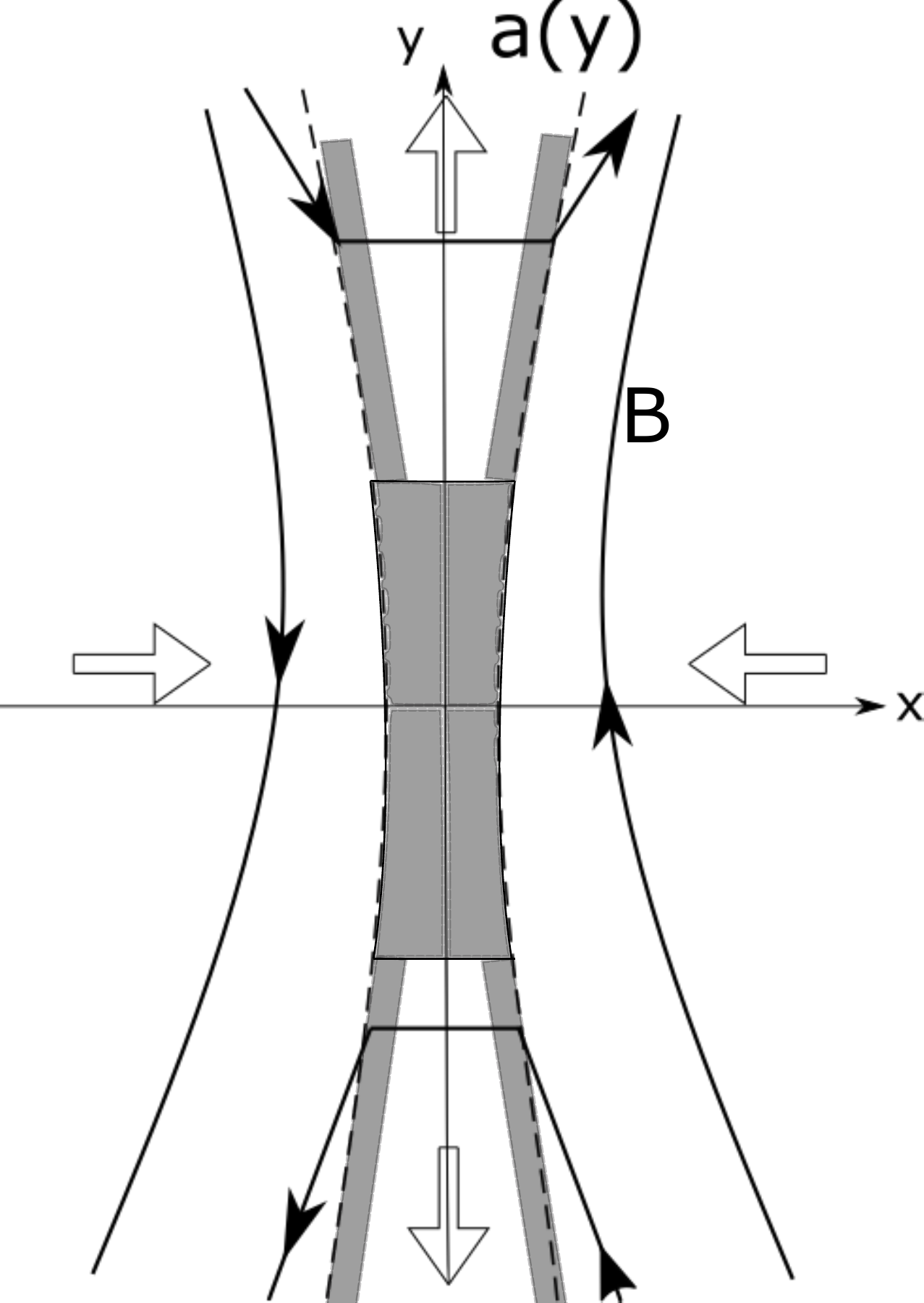}
    \caption{Petschek-type reconnection configuration. The inflow is in the x-direction and the outflow is in the y-direction. Solid arrows indicate magnetic fields. Dashed lines bound the current sheet, inside which are the diffusion region, the transition region, and the downstream region of standing slow shocks.}
    \label{fig:reconnection}
\end{figure*}

To get a more quantitative comparison to our simulation results, here we introduce the averaged MHD equations derived by Refs.~\onlinecite{seaton2009analytical,baty2014formation}. Physical quantities inside the diffusion region, in the steady-state, are averaged across the reconnection layer (illustrated in Fig.~\ref{fig:reconnection}) to reduce the full MHD partial differential equations (PDEs) into a much simpler system of ordinary differential equations (ODEs), that depends only on coordinate y. For a given resistivity profile, one can then solve for the averaged outflow speed, current sheet width, and averaged normal (reconnected) magnetic field. In this work, we apply this theory using the hyperbolic tangent resistivity profiles (Eq.~(\ref{eq:tanh})) and compare the solutions with our numerical simulations in Figures 1, 2, and 4.

The averaged continuity equation, momentum equation, energy equation, and Ohm's law are derived to be

\begin{align}
    &\frac{d}{dy}(a\langle\rho\rangle\langle v_y\rangle)=-\rho_av_{xa},\label{eq:continu}\\
    &\frac{d}{dy}(\langle\rho\rangle\langle v_y\rangle^2a) = -a\frac{d\langle p\rangle}{dy}+B_{ya}\langle B_x\rangle,\label{eq:mom}\\
    &\frac{d}{dy}\bigg\{\left[\frac{\langle\rho\rangle\langle v_y\rangle^2}{2}+\frac{\gamma\langle p\rangle}{\gamma-1}\right]a\langle v_y\rangle\bigg\} = -\left[\frac{\gamma p_a}{\gamma-1}+B_{ya}^2\right]v_{xa},\\
    &E_a = \langle v_y\rangle\langle B_x\rangle+\frac{\eta B_{ya}}{a},\label{eq:ohm}
\end{align}
where $a(y)$ is the half-width of the current sheet, physical quantities with subscript ``$a$'' indicate their values at $x=a(y)$, and the averaged quantities are defined by averaging over x from 0 to $a(y)$,
\begin{align}
    \langle A \rangle(y) \equiv \frac{1}{a}\int_0^a A(x,y)dx.
\end{align}

In the following, magnetic field $B$ is normalized to the upstream magnetic field ($B_0$), velocity to the upstream Alf\'ven speed ($B_0/\sqrt{\mu_0\rho_0})$, density to the density at $a(y)$ ($\rho_a$), pressure to the upstream magnetic pressure ($B_0^2/2\mu_0$), $\mu_0=1$, and length to the system size ($L$). Therefore, $B_{ya}=1$, $\rho_a=1$, $p_a=\beta/2$ are assumed, and we drop off the angle brackets for convenience. The averaged-equations become
\begin{align}
     \label{A6}
    &a\rho v_y = M_A(y-y_{sp})\\ 
     \label{A7}
    &\frac{d}{dy}(\rho v_y^2a) = B_x\\
    \label{A8}
    &\left[\frac{\rho v_y^2}{2}+\frac{\gamma(1+\beta)}{2(\gamma-1)}\right]av_y=\left[\frac{\gamma\beta}{(\gamma-1)2}+1\right]M_A(y-y_{sp})\\
     \label{A9}
    &E_a = M_A = -v_{xa}B_{ya}=  B_xv_y+\frac{\eta(y)}{a},
\end{align}
where $y_{sp}$ is the position of the flow stagnation point and $M_A$ is the inflow Alf\'ven Mach number. Baty et. al.\cite{baty2014formation} further combines these equations into a single ODE by plugging Eqs.~(\ref{A6}), (\ref{A7}) and (\ref{A8}) into Eq.(\ref{A9}) ,
\begin{align}
    (y-y_{sp})\frac{d v_y}{d y}+v_y=\frac{1}{v_y}-\frac{\rho\eta(y)}{M_A^2(y-y_{sp})},\label{eq:vy}
\end{align}
where 
\begin{align}
    \rho = \frac{5(1+\beta)}{5\beta+4-2v_y^2}
\end{align}
is the averaged density of plasma inside the diffusion region and $\gamma=5/3$ is used.

Given an $\eta(y)$ profile, $M_A$ and $y_{sp}$ can be solved by expanding $v_y$ and $\eta$ with respect to $y_{sp}$
\begin{align}
    v_y &= v_1(y-y_{sp})+v_2(y-y_{sp})^2+v_3(y-y_{sp})^3+...\\
    \eta &= \eta_0+\eta_1(y-y_{sp})+\eta_2(y-y_{sp})^2+...,
\end{align}
By plugging these expansions into equation (\ref{eq:vy}) and solving for the coefficients, one can get
\begin{align}
    v_1 = \frac{(5\beta+4)M_A^2}{5(1+\beta)\eta_0}\\
    v_2 = -\frac{\eta_1(5\beta+4)M_A^2}{\eta_0^25(1+\beta)}
\end{align}
\begin{align}
    &v_3 = \frac{(5\beta+4)M_A^2}{25(1+\beta)^2\eta_0^3}\left[-2(4+5\beta)M_A^4+5(1+\beta)(\eta_1^2-\eta_0\eta_2)\right]\label{v3}\\
    &v_4 = \frac{(4+5\beta)M_A^2}{125(1+\beta)^3\eta_0^4}[25 (1 + \beta)^2( -\eta_1^3 + 2 \eta_0 \eta_1 \eta_2 - \eta_0^2 \eta_3) \notag\\
    &+ (4+5\beta)(34+35\beta) \eta_1 M_A^4]\label{v4} 
\end{align}

The condition for convergence is obtained by requiring the coefficients of higher order terms to vanish, \[\lim_{n\rightarrow\infty}v_n=0.\]
We take $v_n=0$ and $v_{n+1}=0$ for $n=3$ since $v_1$ and $v_2$ are the lowest order terms needed to reproduce the bi-directional outflows. Baty et. al.\cite{baty2014formation} solved $M_A$ and $y_{sp}$ using different values of $n$ and showed that their numerical values do not change much if $n\geq 3$, as shown in table \rom{1} and \rom{2} in their paper. After solving numerical values of $M_A$ and $y_{sp}$ by setting Eqs.~(\ref{v3}) and (\ref{v4}) to zeros, the averaged outflow speed $v_y$ can be obtained numerically by solving Eq. (\ref{eq:vy}). Note that the averaged equation Eq. (\ref{eq:ohm}) should capture the physics of Kulsrud's mechanism. To see this, we take the y-derivative of the z-component of the averaged Ohm's law, which gives
\begin{align}
    &\partial_y\int^a_0E_z(x,y)dx=\notag\\
    &\partial_y\int^a_0v_y(x,y)B_x(x,y)dx+\partial_y\int^a_0\eta(x,y) J_z(x,y)dx.
\end{align}
Applying the fundamental theorem of calculus, we get
\begin{align}
    &\int^a_0\partial_yE_z(x,y)dx+\frac{da}{dy}E_z(a,y)=\notag\\
    &\int^a_0\partial_y\left(v_y(x,y)B_x(x,y)\right)dx+\frac{da}{dy}v_y(a,y)B_x(a,y)+\notag\\
    &\int^a_0\partial_y\left(\eta(x,y) J_z(x,y)\right)dx+\frac{da}{dy}\eta(a,y)J_z(a,y).
\end{align}
Using the un-averaged Ohm's law,
\begin{align}
    E_z(a,y)=v_y(a,y)B_x(a,y)+\eta(a,y)J_z(a,y),
\end{align}
we have 
\begin{align}
    &\int^a_0\partial_yE_z(x,y)dx=\notag\\
    &\int^a_0\partial_y\left(v_y(x,y)B_x(x,y)\right)dx+\int^a_0\partial_y\left(\eta(x,y) J_z(x,y)\right)dx,
\end{align}
This is the induction equation (\ref{eq:ind}) averaged over x from 0 to $a(y)$, which was used to derive Eq.~(\ref{Kulsred}). Figure~\ref{fig:1e-3_2e-4} has shown the prediction of $a(y)$ in panel (a), $v_y(y)$ in panel (b) and $B_x(y)$ in panel (c) for Run $T_1$. The agreement is reasonable.


\end{document}